# Fractional Laplacian, Lévy stable distribution, and time-space models for linear and nonlinear frequency-dependent lossy media

W. Chen[a)] (corresponding author) and S. Holm

Simula Research Laboratory

P. O. Box. 134

1325 Lysaker

Oslo, Norway

[a)]E-mail: wenc@simula.no

**Running title**: Fractional Laplacian modeling of lossy media

**Abstract**

The frequency-dependent attenuation typically obeys an empirical power law with an exponent ranging from 0 to 2. The standard time-domain partial differential equation models can describe merely two extreme cases of frequency independent and frequency-squared dependent attenuations. The otherwise non-zero and non-square frequency dependency occurring in many cases of practical interest is thus often called the anomalous attenuation. In this study, we developed a linear integro-differential equation wave model for the anomalous attenuation by using the space fractional Laplacian operation, and the strategy is then extended to the nonlinear Burgers, KZK, and Westervelt equations. A new definition of the fractional Laplacian is also introduced which naturally includes the boundary conditions and has inherent regularization to ease the hyper-singularity in the conventional fractional Laplacian. Under the Szabo's smallness approximation where attenuation is assumed to be much smaller than the wave number, our linear model is found consistent with arbitrary frequency dependencies. According to the fact that the physical attenuation can be understood a statistic process, the empirical range [0,2] of the power law exponent is explained via the Lévy stable distribution theory. It is noted that the power law attenuation underlies fractal microstructures of anomalously attenuative media.

PACS numbers: 43.20.Bi, 43.20.Hq, 43.35.Bf, 43.35.Cg

**INTRODUCTION**

The frequency-dependent attenuation has been observed in a wide range of important engineering areas such as acoustics (Blackstock 1985, Szabo, 1994; Wojcik et al, 1995), viscous dampers in seismic isolation of buildings (Makris and Constantinou, 1991), structural vibration (Enelund, 1996; Rusovici, 1999; Adhikari, 2000), seismic wave propagation (Caputo, 1967; Caputo and Mainardi, 1971), anomalous diffusions occurring in porous media (Hanyga, 1999), just to mention a few. This frequency dependency is described by

$$E = E_0 e^{-\alpha(\omega)z}, \tag{1}$$

where $E$ denotes the amplitude of an acoustic field variable such as velocity or pressure, and $\omega$ represents angular frequency. Coefficient $\alpha(\omega)$ is often characterized with an empirical power law

$$\alpha(\omega) = \alpha_0 |\omega|^y, \qquad y \in [0,2], \tag{2}$$

for a wide range of frequencies of practical interest, in which $\alpha_0$ and $y$ are media-specific attenuation parameters obtained through a fitting of measured data.

The most straightforward strategy in computer simulation of the power law lossy behavior is to do both mathematical and numerical modeling in the frequency domain via the Laplace transform (Ginter, 2000). The drawbacks of this approach are that the frequency domain methods are often ineffective for nonlinear problems and the numerical inverse Laplace transform is very tedious and expensive. The time-domain simulation, in contrast, is feasible for general nonlinear problems and relatively easier to implement and less costly (Wismer and Ludwig, 1995). In addition, the time domain models also outperform the frequency domain models as they allow numerical simulation of various initial and boundary value problems (Hanyga, 2001a).

However, it has long been noted that common time-domain partial differential equations (PDE) can model merely two extreme cases of frequency-independent ($y=0$) and frequency-squared dependent ($y=2$) absorption behaviors. In many cases of practical interest such as acoustics in biomedical materials and fractal rock layers, $0<y<2$ mostly appears and the standard time-domain PDE modeling methodology does not apply (Blackstock, 1985; Nachman et al, 1990; Szabo, 1994). In contrast to the $y=0,2$ attenuations well described by the standard PDEs, the attenuations obeying $0<y<2$ power law are thus often called the anomalous diffusion (Hanyga, 2001c), non-exponential relaxation, inelastic damping (Adhikari, 2000), hysteretic damping (Gaul, 1999), singular hereditary or singular memory media (Hanyga, 1999), originating from different engineering applications. The recent decade has witnessed increasing attention to accurate time-domain mathematical modelling of such anomalous ($0<y<2$) attenuation phenomena due to a dramatic increase in computer simulation of acoustic wave propagation through human tissues and irregular porous random media.

There are actually a few time-domain models available today to tackle the mathematical difficulty of modeling attenuation anomaly ($0<y<2$) in different ways. A simple approach is to use the adaptive Rayleigh proportional damping model (Wojcik et al, 1995, 1999). The classic Rayleigh viscous model is essentially stated as

$$\alpha(\omega) = \beta_1 + \beta_2 \omega^2, \qquad (3)$$

where $\beta_1$ and $\beta_2$ are two Rayleigh coefficients. $\beta_1=0$ means attenuation will increase quadratically with frequency, while $\beta_2=0$ leads to a frequency-independent model. The classic Rayleigh model is therefore considered to exactly reproduce only zero and quadratic frequency dependences and has the corresponding normal PDE models. By carefully manipulating both $\beta_1$ and $\beta_2$, the basic idea behind the adaptive Rayleigh model is to fit

$$\alpha_0 \omega^y = \beta_1 + \beta_2 \omega^2 \qquad (4)$$

in terms of media-specific exponent *y* ranging between 0 and 2. In fact, this strategy models an anomalous attenuation via a weighted sum of the standard PDE models of frequency-independent and frequency-squared dependent attenuations, which, under the smallness approximation (Szabo, 1994), can be formulated as

$$\Delta p = \frac{1}{c_0^2}\frac{\partial^2 p}{\partial t^2} + \frac{\partial}{\partial t}(\beta_1 - \beta_2 \Delta)p, \qquad (5)$$

where $\Delta$ denotes the standard Laplace operator. For more related details on (5) see the thermoviscous equation (22) and the damped wave equation (25) in the later section (3). The adaptive Rayleigh approach is very simple to implement in the standard numerical codes, computationally efficient, and thus attractive in engineering simulation. The major drawbacks are that the determination of $\beta_1$ and $\beta_2$ is somewhat tricky and it is difficult to accurately fit formula (4) over a large range of frequencies, which typically takes place in broadband pulse propagations.

By using a convolution integral of the physical signal time derivative and the medium compressibility, Nachman et al. (1990) introduced a multiple relaxation model to capture arbitrary power law dependence on frequency without introducing of nonphysical dispersion or violation of causality (Mast et al, 2001). For the N-th order relaxation model, the medium compressibility $\kappa$ is expressed as

$$\kappa(\bar{x},t) = \kappa_\infty(\bar{x})\delta(t) + \sum_{i=1}^{N}\frac{\kappa_i(\bar{x})}{\tau_i(\bar{x})}e^{-t/\tau_i}H(t), \qquad (6)$$

where $\kappa_\infty$ denotes the compressibility at infinite frequency; $\tau_i$ and $\kappa_i$ respectively represent the i-th order relaxation time and relaxation modulus; $\delta$ is the Dirac delta function, $H(t)$ the Heaviside function to guarantee causality, and $\bar{x}$ the location coordinate. This multi-relaxation model has been successfully tested in numerical simulation of lossy broadband acoustic wave propagations (Yuan et al, 1999), and appears to work well for the relaxation-dominated attenuation. However, the direct numerical simulation of this model entails very expensive time convolution computation. Yuan et al (1999) developed a sophisticated numerical implementation scheme for this model to improve computational efficiency. The multiple compressibility expression (6) also shows that the N-th order multi-relaxation model requires tailoring *N* relaxation times and compressibility parameters to fit problem-dependent $\alpha_0$ and *y* of the empirical power law formula (2) over a range of frequencies. This could be a quite tricky and tedious task. In addition, as shown in the fractional calculus models discussed below, the singular part of the convolution kernel plays a central role for many memory media and cannot well be approximated by a discrete set of relaxation mechanisms of the Maxwell-Voigt or Zener type (Hanyga, 1999). In other words, fitting via relaxation constants over a certain bandwidth does not guarantee an exact description of anomalous attenuation.

On the other hand, the time fractional derivative has long been found to be a very effective means to describe the anomalous attenuation behaviors. For instance, Caputo (1967) and Caputo and Mainardi (1971) developed the constant *Q* seismic wave propagation model via the fractional derivative, where *Q* represents an attenuation per unit wavelength. The Bagley-Torvik attenuation model (Bagley and Torvik, 1983) for the range 1<*y*<2 has also been widely noticed, where the fractional derivative is an essential building block. A fractional derivative time model without space variables (Seredynska and Hanyga, 2000) can typically be expressed as

$$D^2 u + \gamma D^{1+\eta} u + F(u) = 0, \qquad (7)$$

where $D^{1+\eta}$ with 0<$\eta$<1 represents the (1+$\eta$)-order fractional derivative in the sense of Caputo (1967) and models the anomalous attenuation, and $\gamma$ is the thermoviscous coefficient. The detailed definitions of the fractional derivative will be given in later section 2. Gaul (1999) points out that replacing the integer time derivative in the conventional PDE attenuation modelling (e.g., the preceding adaptive Rayleigh proportional and multi-relaxation models) by the fractional time derivative (e.g., $D^{1+\eta} u$ in (7)) leads to better accuracy with fewer parameters and guaranteed causality. Note that unlike the derivatives of the integer order, the derivatives of the fractional order are non-local and involve the whole history of the function evolution, and can effectively describe physical memory behaviors. The uniqueness, existence and well-posedness of the fractional derivative equations have been thoroughly researched and assured (Hanyga, 2001b).

As a new alternative methodology to the fractional derivative, Szabo (1993, 1994) and Szabo and Wu (2000) introduced the constitutive attenuation models with singular convolution kernels along with the time-domain causal dispersion-attenuation relations instead of the frequency domain Kramers-Kronig relations. Szabo's models directly use the parameters $\alpha_0$ and $y$ to be fully consistent with the empirical frequency-dependent attenuation power law. One of the shortcomings of Szabo's models is that they confront hyper-singular convolution integral kernels. For example, the Szabo's linear lossy wave model for the non-integer exponent $y$ is given by

$$\Delta p = \frac{1}{c_0^2}\frac{\partial^2 p}{\partial t^2} + \frac{4\alpha_0 \Gamma(y+2)\cos((y+1)\pi/2)}{\pi c_0}\int_{-\infty}^{t}\frac{p(\tau)}{(t-\tau)^{y+2}}d\tau, \qquad (8)$$

where $p$ represents the pressure; and $\Gamma$ is the Euler's gamma function. The second left-hand term of attenuation is an improper hyper-singular integral. It is noted that the convolution integral appears in all the preceding time-domain models but the adaptive Rayleigh model. Therefore, the models of the convolution integral type are often called the pseudo-differential equations (Gorenflo and Mainardi, 1998) or the integro-differential equation (Hanyga, 2001a) rather than the partial differential equation.

The hyper-singular integral inside (8) must be interpreted in the sense of the Hadamard finite-part integral, and its numerical solution is still an open research topic. As mentioned by Blackstock (1985), the space or space-time modelling of thermoviscous behavior is often replaced by a pure time operation under the condition that the thermoviscous term is relatively small. Ochmann and Makarov (1993) further elaborate that this replacement is impossible in the general case where the interaction between two oppositely traveling sound waves can not be neglected. The preference of the time-only expression is mostly due to its ease of analysis. For instance, the time-space representation $\partial^3 p/\partial t \partial z^2$ in the one-dimension thermoviscous wave equation, where *t* and *z* are respectively time and space variables (see equation (26)), is approximated by a triple time derivative $\partial^3 p/\partial t^3$ (Blackstock 1967; Pierce, 1989; Szabo, 1994). However, numerical implementations of the time-only models are still uncommon, and most research is now restricted to the related mathematical analysis partly due to great numerical difficulties involved. In addition, when 2<1+$\eta$<3, fractional time derivative $D^{1+\eta}u$ in (7) involves the initial condition of the second order derivative which is unavailable in most practical problems. It is well known that anomalously attenuative and dispersive media often establish complicated microstructures in space, the spatial fractional derivative models may therefore instead be more suitable as a modelling approach, where the initial condition of second order derivative is never required (Hanyga, 2001b).

The purpose of this study is to employ the spatial fractional Laplacian, also sometimes called the fractional Laplace operator, instead of the Szabo's time convolutional integral and the time fractional derivative to develop linear and nonlinear mathematical models of anomalous thermoviscous behavior characterized by non-zero and non-quadratic frequency dependency. It is known (Samko et al, 1987) that the standard definition of the fractional Laplacian leads to a hyper-singular convolution integral as in the Riemann-Liouville fractional derivative. We present a new definition of the fractional Laplacian which naturally includes the boundary conditions and has inherent regularization operation to ease the hyper-singularity of the convolution kernel function. Therefore, it is more useful for engineering modelling.

In terms of the Lévy stable distribution theory, the second goal of this research is to provide a physical ground why the exponent *y* of the frequency power law dependence ranges between 0 and 2. It is known that the Gaussian statistic Brownian motion is related to normal diffusion, while statistically speaking, the non-Gaussian process corresponds to anomalous diffusion which can be described by the Lévy $\beta$-stable distribution theory, where the stability index $\beta$ is bounded within 0<$\beta$≤2. We find that both parameters *y* and $\beta$ are inherently interconnected thorough an analysis of the equivalence of the probability density function of the Lévy stable distribution and the fundamental solution of the presented fractional Laplacian model equations.

In what follows, the new definition of the fractional Laplacian is introduced first in section I, followed by a presentation and analysis of the linear fractional Laplacian thermoviscous models of wave equation in section II. The corresponding nonlinear models are developed in section III. Section IV is dedicated to the physical explanation of the [0,2] range of exponent *y* via the Lévy stable distribution theory. The conclusions are found in section V based on the results reported herein. Finally, the appendix briefly discusses the connection between power law attenuation, fractional Laplacian, Lévy distribution, and fractals of media microstructures.

## I. FRACTIONAL LAPLACIAN

It is worth pointing out that the fractional Laplacian and the fractional derivative are two related but different mathematical concepts. Both are defined through a singular convolution integral, but the former is guaranteed to be the positive definition like the standard Laplace operator, while the latter is not (Diethelm, 2000).

The conventions of Fourier transform used in Pierce (1989) and Szabo (1994a) are also employed in this study. Namely,

$$F_{-}\left(\frac{\partial^{n}\phi}{\partial z^{n}}\right)=(ik)^{n}\Phi(k,\omega), \qquad (9)$$

$$F_{+}\left(\frac{\partial^{n}\phi}{\partial t^{n}}\right)=(-i\omega)^{n}\Phi(k,\omega), \qquad (10)$$

where $\Phi(k,\omega)$ is the time and space two dimensional Fourier transforms of a sufficiently good function $\phi(z,t)$,

$$\Phi(k,\omega) = \int_{-\infty}^{\infty}\int_{-\infty}^{\infty} \phi(z,t)e^{-i(kz-\omega t)}dzdt. \tag{11}$$

$k$ is the wave number. The inverse of the space Fourier transform is designated as $F_-^{-1}$, and the inverse of the time Fourier transform $F_+^{-1}$.

A common interpretation of the fractional Laplacian is to employ the inverse of its Fourier transform (e.g., see Samko et al, 1987; Jespersen, 1999), i.e.,

$$F_-\left\{(-\Delta)_*^{s/2}\varphi\right\} = k^s\Phi, \qquad 0<s<2, \tag{12}$$

$$(-\Delta)_*^{s/2}\varphi = F_-^{-1}\left\{k^s\Phi\right\} = \frac{1}{2\pi}\int \Phi k^s e^{ikx}dk. \tag{13}$$

The fractional Laplacian is also often called the Riesz fractional derivative in terms of the Riesz potential (Gorenflo and Mainardi, 1998). The Riesz potential $I_d^s$ of order $s$ of $n$ dimensions reads (Zahle, 1997; Samko et al, 1987)

$$I_d^s\varphi(x) = \frac{\Gamma[(d-s)/2]}{\pi^{s/2}2^s\Gamma(s/2)}\int_\Omega \frac{\varphi(\xi)}{\|x-\xi\|^{d-s}}d\Omega(\xi), \quad 0<s<2, \tag{14}$$

where $\Gamma$ denotes the Euler's gamma function, $\Omega$ is integral domain, and d dimension. The definition of the fractional Laplacian can be stated as

$$(-\Delta)_*^{s/2}\varphi(x) = -\Delta\left[I_d^{2-s}\varphi(x)\right]. \tag{15}$$

It is known that the radial Laplacian operator has the expression

$$\Delta \varphi(x) = \frac{d^2 \varphi}{dr^2} + \frac{d-1}{r}\frac{d\varphi}{dr}, \tag{16}$$

where $r = \|x - \xi\|$. (15) can then be reduced to

$$\begin{aligned}(-\Delta)_*^{s/2} \varphi(x) &= -\frac{\Gamma[(d-2+s)/2]}{\pi^{2-s/2} 2^{2-s} \Gamma[(2-s)/2]} \Delta \int_\Omega \frac{\varphi(\xi)}{\|x-\xi\|^{d-2+s}} d\Omega(\xi) \\ &= -\frac{(d-2+s)s\Gamma[(d-2+s)/2]}{\pi^{(2-s)/2} 2^{2-s} \Gamma[(2-s)/2]} \int_\Omega \frac{\varphi(\xi)}{\|x-\xi\|^{d+s}} d\Omega(\xi).\end{aligned} \tag{17}$$

It is noted that (17) encounters the detrimental issues such as the hyper-singularity. An alternative way is thus presented below to define the fractional Laplacian without the perplexing hypersingular issues in the Riesz fractional derivative (15)

$$\begin{aligned}(-\Delta)^{s/2} \varphi(x) &= -I_d^{2-s}[\Delta \varphi(x)] \\ &= -\frac{\Gamma[(d-2+s)/2]}{\pi^{(2-s)/2} 2^{2-s} \Gamma[(2-s)/2]} \int_\Omega \frac{\Delta \varphi(\xi)}{\|x-\xi\|^{d-2+s}} d\Omega(\xi).\end{aligned} \tag{18}$$

The Green second identity is useful to simplify (18) and can be stated as

$$\int_\Omega v \Delta \varphi d\xi = \int_\Omega \varphi \Delta v d\Omega(\xi) - \int_S \left( u \frac{\partial v}{\partial n} - v \frac{\partial \varphi}{\partial n} \right) dS(\xi), \tag{19}$$

where $S$ represents the surface of the domain, and $n$ is the unit outward normal. Let

$$v = 1/\|x-\xi\|^{d-2+s}, \tag{20}$$

and boundary conditions

$$\varphi(x)\big|_{x \in \Gamma_D} = D(x), \tag{21}$$

$$\left.\frac{\partial \varphi(x)}{\partial n}\right|_{x\in \Gamma_N} = N(x), \qquad (22)$$

where $\Gamma_D$ and $\Gamma_N$ are the surface part corresponding to the Dirichlet boundary and the Neumann boundary, and using the Green second identity, the definition (18) is then reduced to

$$(-\Delta)^{s/2}\varphi(x) = -\frac{1}{h}\left\{ \int_\Omega \frac{\varphi(\xi)}{\|x-\xi\|^{d+s}} d\Omega(\xi) - \right.$$

$$\left. \int_S \left[ \varphi(\xi)\frac{\partial}{\partial n}\left(\frac{1}{\|x-\xi\|^{d+s}}\right) - \frac{1}{\|x-\xi\|^{d+s}}\frac{\partial \varphi(\xi)}{\partial n}\right] dS(\xi) \right\} \qquad (23)$$

$$= (-\Delta)_*^{s/2}\varphi(x) + \frac{1}{h}\int_S \left[ D(\xi)\frac{\partial}{\partial n}\left(\frac{1}{\|x-\xi\|^{d+s}}\right) - \frac{N(\xi)}{\|x-\xi\|^{d+s}}\right] dS(\xi),$$

where

$$h = \frac{\pi^{(2-s)/2} 2^{2-s} \Gamma[(2-s)/2]}{(d-2+s)s\Gamma[(d-2+s)/2]}. \qquad (24)$$

It is seen from (23) that the presented fractional Laplacian definition $(-\Delta)^{s/2}$ is thus considered the Riesz fractional derivative $(-\Delta)_*^{s/2}$ (the standard fractional Laplacian) augmented with the boundary integral, which is a parallel to the fractional derivatives in the Caputo sense relative to that in the Riemann-Liouville sense. Comparing the traditional Riesz fractional derivative $(-\Delta)_*^{s/2}$ definition (17) and the present fractional Laplacian $(-\Delta)^{s/2}$ definition (18), one can see that the singularity $d$-2+$s$ of $(-\Delta)^{s/2}$ is smaller than dimension $d$, and thus, $(-\Delta)^{s/2}$ avoids the hyper-singularity of $(-\Delta)_*^{s/2}$.

The above two definitions $(-\Delta)_*^{s/2}$ and $(-\Delta)^{s/2}$ involve only the symmetric fractional Laplacian for isotropic media. To simplify the illustration of the basic idea of this study without loss of generality, we only consider isotropic media in this paper. For the traditional definition of the anisotropic fractional Laplacian see Feller (1971) and Hanyga (2001). By analogy with the new definition (18) and (23), it will be straightforward to have the corresponding new expression of the anisotropic fractional Laplacian.

Albeit a long history, the research on the space fractional Laplacian still appears poor in the literature (Gorenflo and Mainardi, 1998). In recent years, some interest arises from anomalous diffusion problems. The readers are advised to find more detailed description of the fractional Laplacian from Samko et al (1987), Zaslavsky (1994), Gorenflo and Mainardi (1998), Hanyga (2001) and references therein.

## II. LINEAR FRACTIONAL LAPLACIAN THERMOVISCOUS MODEL

Szabo (1994) started his time-domain model building and causality analysis of the attenuation power law with the thermoviscous wave equation (Blackstock, 1967; Lighthill, 1980; Pierce, 1989), also known as the augmented wave equation (Johnson and Dudgeon, 1993), which governs the propagation of sound through a viscous fluid and can be stated as

$$\Delta p = \frac{1}{c_0^2}\frac{\partial^2 p}{\partial t^2} + \frac{\mu}{c_0^2}\frac{\partial}{\partial t}(-\Delta p), \qquad (25)$$

where $c_0$ is inviscid phase velocity, and $\mu = [4\eta/3 + \kappa(\gamma_h - 1)/c_p]/\rho_0$ the collective thermoviscous coefficient, $\eta$ the shear viscosity coefficient, $\rho$ the ambient density, $\kappa$ thermal conductivity, $\gamma_h$ ratio of specific heats, and $c_p$ special heat at constant pressure. Eq. (25) describes both dispersion (waveform alternation with respect to frequency) and attenuation behaviors. Szabo (1994) pointed out that the low-frequency approximation of (25) leads to a square dependence of attenuation on frequency with constants

$$\alpha_0 = \mu/2c_0^3, \quad y=2. \tag{26}$$

Szabo (1994) noted that the exponent of the attenuation, *y*, and the order of the loss term in the generalized wave equation are related. The relationship is that the time differentiation order of the loss term is higher than *y* by one. By analogy with this relationship in generalizing (25) via the space fractional Laplacian, we intuitively get

$$\Delta p = \frac{1}{c_0^2}\frac{\partial^2 p}{\partial t^2} + \frac{2\alpha_0}{c_0^{1-y}}\frac{\partial}{\partial t}(-\Delta)^{y/2} p, \tag{27}$$

where $(-\Delta)^{y/2}$ is the fractional Laplace and 0<*y*<2. The fractional Laplacian is a positive definite operator (Gorenflo and Francesco, 1998) and the wave equation (27) is therefore causal. It is also well known that the Duhamel's principle applies to the fractional Laplacian equations and the causality is guaranteed (Hanyga, 2001c). When *y*=2, (27) is equivalent to the model equation (25). When *y*=0, (27) is reduced to the standard damped wave equation

$$\Delta p = \frac{1}{c_0^2}\frac{\partial^2 p}{\partial t^2} + \frac{2\alpha_0}{c_0}\frac{\partial p}{\partial t}. \tag{28}$$

(28) describes the frequency-independent attenuation (Szabo, 1994).

To facilitate the analysis without loss of generality, let us consider the 1D case of model equation (27)

$$\Delta p = \frac{1}{c_0^2}\frac{\partial^2 p}{\partial t^2} + \frac{2\alpha_0}{c_0^{1-y}}\frac{\partial}{\partial t}\left(-\frac{\partial^2 p}{\partial z^2}\right)^{y/2} p. \tag{29}$$

The Fourier transforms of the time and space derivatives are given below by

$$F_-\left\{\frac{\partial^2 p}{\partial x^2}\right\} = (ik)^2 P = -k^2 P, \tag{30}$$

$$F_-\left\{\left(-\frac{\partial^2}{\partial x^2}\right)^{y/2} p\right\} = \left(-(ik)^2\right)^{y/2} P = k^y P, \tag{31}$$

$$F_+\left\{\frac{\partial^2 p}{\partial t^2}\right\} = (-i\omega)^2 P = -\omega^2 P. \tag{32}$$

Applying the time and space Fourier transforms (30-32) to (29), we have the dispersion equation

$$k^2 - \omega^2/c_0^2 - i\,2\alpha_0\omega k^y/c_0^{1-y} = 0. \tag{33}$$

Since $k = \beta + i\alpha$ and $\beta = \omega/c_p$, it is straightforward to have

$$\beta^2 - \alpha^2 - \omega^2/c_0^2 + i2\alpha\beta - i\,2\alpha_0\omega\beta^y(1+i\alpha/\beta)^y/c_0^{1-y} = 0. \tag{34}$$

To move the analysis further, the Szabo conservative smallness approximation (equations (24) and (25) of Szabo (1994)) is crucial, i.e.,

$$\alpha/\beta \approx \alpha_0|\omega|_{\lim}^y/\beta_0 = \alpha_0|\omega|_{\lim}^{y-1}c_0 \leq 0.1, \tag{35}$$

where $\omega_{\lim}$ is the frequency limit corresponding to 0.1. As discussed in Szabo (1994), the limit frequency range in terms of (35) is adequate enough to cover the frequency spectrum of practical interest in medical ultrasound applications. In terms of (35), (34) is then approximated by the binomial expansion as

$$\beta^2 - \alpha^2 - \omega^2/c_0^2 + i2\alpha\beta - i2\alpha_0\omega\beta^y/c_0^{1-y} + 2\alpha_0 y\omega\alpha\beta^{y-1}/c_0^{1-y} = 0. \quad (36)$$

With the further help of the smallness approximation (35) and $\beta_0 = \omega/c_0$, from (36) we derive

$$\alpha \approx \alpha_0 |\omega|^y, \quad (37a)$$

$$\beta \approx \beta_0 = \omega/c_0. \quad (37b)$$

It is noted that (37a) matches the power law (2). By now we have shown that the fractional Laplacian model (29) does have a power law attenuation under the Szabo smallness approximation condition (35) corresponding to the time convolutional integral model (Szabo, 1994) and the time fractional models (Baglegy and Torvik, 1983).

**III. NONLINEAR LOSSY MEDIA**

Most nonlinear acoustic equations are only useful for describing lossy media with a quadratic dependence or independence on frequency (Szabo, 1993), and thus have limited practical utility. In this section, we will extend the fractional Laplacian modelling of anomalous thermoviscous effect to some well known nonlinear acoustics equations for accommodating lossy media obeying the power law of arbitrary exponents. It is noted that the thermoviscous representation in the standard nonlinear acoustic PDE models, which characterizes the effects of absorption and dispersion, is mostly the same as in the corresponding linear models. Thus, Blackstock (1985) presented a straightforward strategy constructing the nonlinear anomalous attenuation model by simply replacing its attenuation term with that in the corresponding linear model, while keeping all other linear and nonlinear terms unchanged. The methodology is justified by a perturbation analysis (Blackstock, 1985). Following this strategy, Szabo (1993) extended his linear convolution integral modelling of arbitrary power law exponents (Szabo, 1994) to the Burgers, KZK, and Westervelt equations. By analogy with Blackstock (1985) and Szabo (1993), this study generalizes these nonlinear acoustic models by replacing their thermoviscous term with the fractional Laplacian lossy terms given in equations (27) and (29).

It is stressed that the smallness approximation condition (35), crucial in the analysis of the preceding linear fractional Laplacian models, is also the foundation of the nonlinear acoustics modelling (Hamilton and Blackstock, 1998; Szabo, 1993). Thus, all derivations here are consistent.

**A. Fractional Laplacian Burgers equation**

The Burgers equation may be the best-known simple nonlinear acoustic model which describes the combined effects of nonlinearity and dissipation. The one-dimensional Burgers equation for plane progressive waves is stated as

$$\frac{\partial p}{\partial t} + \beta p \frac{\partial p}{\partial z} - \varepsilon \frac{\partial^2 p}{\partial z^2} = 0, \tag{38}$$

where $\beta$ denotes the nonlinearity coefficient; and $\varepsilon$ is a constant proportional to the thermoviscous absorption coefficient. It is known (Blackstock 1985, Szabo, 1993) that the Burgers equation (38) describes lossy acoustic propagation of square frequency dependence. To extend the Burgers equation to reflect power law media of arbitrary exponent $y$, Blackstock (1985) suggests and verifies to some extent that only the second term of (38) needs to be modified which involves the absorption, while keeping all others the same.

As detailed by Szabo (1994), the hyperbolic wave equation can be approximated to the parabolic equation by removing the left-hand side term of (29), namely,

$$\frac{1}{c_0^2}\frac{\partial^2 p}{\partial t^2} + \frac{2\alpha_0}{c_0^{1-y}}\frac{\partial}{\partial t}\left(-\frac{\partial^2}{\partial z^2}\right)^{y/2} p = 0. \tag{39}$$

And then, integrating (39) with respect to time $t$ and multiplying by $c_0^2$, we have

$$\frac{\partial p}{\partial t} + 2\alpha_0 c_0^{1+y}\left(-\frac{\partial^2}{\partial z^2}\right)^{y/2} p = 0. \tag{40}$$

The model (40) is a generalized diffusion equation and corresponds to the Burgers equation (38) without the nonlinear convection term. The analytical solution of equation (40) can be found in Hanyga (2001c). Applying the spatial Fourier transform (31) to (40), we have

$$\frac{dP}{dt} + 2\alpha_0 c_0^{1+y} k^{y/2} P = 0. \tag{41}$$

Thus, the transformation solution is

$$P(k,t) = Ce^{-2\alpha_0 c_0^{1+y} t k^y}, \tag{42}$$

where $C$ depends on the initial condition. When $y=2$, (42) exhibits the normal frequency-squared diffusion. It is clear that Eq. (40) is a parabolic model originating from equation (29) while holding the capability describing arbitrary power ($y$) law attenuation. In the following section 5, we will give a further analysis of (42) in relation to the Lévy stable distribution.

By analogy with the generalizing methodology presented by Blackstock (1985) and Szabo (1993), the fractional Laplacian Burgers equation is presented below by simply adding the nonlinear convection term of Burgers equation (38) to Eq. (40)

$$\frac{\partial p}{\partial t} + \beta p \frac{\partial p}{\partial x} + 2\alpha_0 c_0^{1+y}\left(-\frac{\partial^2}{\partial z^2}\right)^{y/2} p = 0. \tag{43}$$

Note that the nonlinear term in (43) can be considered the source term in the sense of an inhomogeneous equation (Szabo, 1993).

The nonlinear equation model (43) belongs to the so-called fractal Burgers equations or the fractional advection-dispersion equation (Schumer et al., 2002). A detailed analysis of such equations is given in Biler et al (2001). In higher dimensional cases, (43) is restated as

$$\frac{\partial p}{\partial t} + \beta p \cdot \nabla p + 2\alpha_0 c_0^{1+y}(-\Delta)^{y/2} p = 0, \qquad (44)$$

where $\nabla p$ represents the pressure gradient vector, and the dot stands for a scalar product. Ochmann and Makarov (1993) also developed the time fractional derivative Burgers equation to describe the power law absorptions with arbitrary $y$.

**B. Fractional Laplacian KZK equation**

Besides absorption and nonlinearity effects included in the Burgers equation, the KZK equation also accounts for the diffraction effect. As a penalty of this more accurate modelling, the KZK equation is also more complicated than the Burgers equation and is expressed as (Hamilton and Blackstock, 1998)

$$\frac{\partial^2 \hat{p}}{\partial z \partial \tau} - \frac{c_0}{2}\nabla_\perp^2 \hat{p} - \frac{\delta}{2c_0^3}\frac{\partial^3 \hat{p}}{\partial \tau^3} = \frac{B}{\rho_0 c_0^3}\hat{p}\frac{\partial \hat{p}}{\partial \tau}, \qquad (45)$$

where the retarded time $\tau = t - z/c_0$, $\nabla_\perp^2 = \partial^2/\partial x^2 + \partial^2/\partial y^2$ is the transverse components of the Laplacian, and the hatted $\hat{p}$ denotes the pressure dependence on $\tau$ compared with $p(t)$. The model (45) is still a parabolic wave equation, which differs from the Burgers equation (38) in the addition of the second left-hand term which represents the planar diffraction effects of the geometrical features, important in high frequency ultrasound beams. The limits to applying the KZK model are 1) that the beam needs to be reasonably directional, where the near and far fields have respectively approximate planar and spherical wavefronts; 2) the ratio of source dimension & wavelength is greater than 0.1; 3) the region of interest is far away from source (Hamilton and Blackstock, 1998).

In terms of the retarded time variable $\tau$, the linear fractional Laplacian loss model (29) is written as

$$\frac{\partial^2 \hat{p}}{\partial z^2} - \frac{2}{c_0}\frac{\partial^2 \hat{p}}{\partial z \partial \tau} - \frac{2\alpha_0}{c_0^{1-y}}\frac{\partial}{\partial \tau}\left(-\frac{\partial^2}{\partial z^2}\right)^{y/2} \hat{p} = 0. \qquad (46)$$

Following the parabolic processing detailed in Szabo (1994), we remove the first Laplacian term under the assumption that it is relatively small, multiply (46) by $c_0/2$ and have

$$\frac{\partial^2 \hat{p}}{\partial z \partial \tau} + \alpha_0 c_0^y \frac{\partial}{\partial \tau}\left(-\frac{\partial^2}{\partial z^2}\right)^{y/2} \hat{p} = 0. \qquad (47)$$

Integrating (47) with respect to $\tau$ leads to a parabolic equation corresponding to the model (29)

$$\frac{\partial \hat{p}}{\partial z} + \alpha_0 c_0^y \left(-\frac{\partial^2}{\partial z^2}\right)^{y/2} \hat{p} = 0. \qquad (48)$$

As done previously with the fractional Laplacian Burgers equation, the fractional Laplacian KZK equation is given by simply replacing the linear thermoviscous term of (45) with the new one given in (47)

$$\frac{\partial^2 \hat{p}}{\partial z \partial \tau} - \frac{c_0}{2}\Delta_\perp \hat{p} + \alpha_0 c_0^y \frac{\partial}{\partial \tau}\left(-\frac{\partial^2}{\partial z^2}\right)^{y/2} \hat{p} = \frac{B}{\rho_0 c_0^3} \hat{p} \frac{\partial \hat{p}}{\partial \tau}. \qquad (49)$$

(49) can apply to nonlinear lossy media following a power law attenuation with arbitrary exponent $y$.

**C. Fractional Laplacian Westervelt equation**

The well-known Westervelt equation reads

$$\left(\Delta - \frac{1}{c_0^2}\frac{\partial^2}{\partial t^2}\right)p + \frac{\delta}{c_0^4}\frac{\partial^3 p}{\partial t^3} = -\frac{B}{\rho_0 c_0^4}\frac{\partial^2 p^2}{\partial t^2}. \tag{50}$$

The model is applicable when cumulative nonlinear effects dominate over local nonlinear effects (Hamilton and Blackstock, 1998). In the cases of compound wave propagation such as standing waves or guided waves in nonplanar models, which encounter significant local effects, the Westervelt model does not hold.

As such, it is straightforward to get the fractional Laplacian Westervelt equation

$$\left(\Delta - \frac{1}{c_0^2}\frac{\partial^2}{\partial t^2}\right)p - \frac{2\alpha_0}{c_0^{1-y}}\frac{\partial}{\partial t}(-\Delta)^{y/2} p = -\frac{B}{\rho_0 c_0^4}\frac{\partial^2 p^2}{\partial t^2}. \tag{51}$$

in terms of the linear fractional Laplacian model (27).

**D. General second-order approximation model**

The Westervelt model is the most sophisticated model discussed up to now. Even so, the model considers only longitudinal nonlinearity rather than transverse nonlinearity, among other restrains mentioned previously. The Burgers, KZK, and Westervelt models are in fact simplifications of varying degree of the following general second-order approximation model

$$\left(\nabla^2 - \frac{1}{c_0^2}\frac{\partial^2}{\partial t^2}\right)p + \frac{\delta}{c_0^4}\frac{\partial^3 p}{\partial t^3} = -\frac{B}{\rho_0 c_0^4}\frac{\partial^2 p^2}{\partial t^2} - \left(\nabla^2 + \frac{1}{c_0^2}\frac{\partial^2}{\partial t^2}\right)L, \tag{52}$$

where $L = \frac{1}{2}\rho_0 u^2 - \frac{p^2}{2\rho_0 c_0^2}$ is the second order Lagrangian density indicating local nonlinear effects. The second-order approximation means that the model (52) drops small terms of higher than the second order. For details see chapter 3 of Hamilton and Blackstock (1998). Albeit much more complex, the model (52) is derived under constraints of suitable small Mach number and does not work well for large amplitudes.

It is noted that only the linear thermoviscous term is included in (52). Along the line leading to the foregoing fractional Laplacian Burgers, KZK, and Westervelt equations, we replace the normal thermoviscous term in (52) by the new loss and dispersion terms presented in linear model (27) and then have the fractional Laplacian nonlinear acoustic equation of general second-order approximation

$$\left(\nabla^2 - \frac{1}{c_0^2}\frac{\partial^2}{\partial t^2}\right)p - \frac{2\alpha_0}{c_0^{1-y}}\frac{\partial}{\partial t}(-\Delta)^{y/2} p = -\frac{B}{\rho_0 c_0^4}\frac{\partial^2 p^2}{\partial t^2} - \left(\nabla^2 + \frac{1}{c_0^2}\frac{\partial^2}{\partial t^2}\right)L. \quad (53)$$

## IV. $\beta$-STABLE LEVY PROCESSS AND EXPONENT RANGE 0<y ≤2 OF POWER LAW ATTENUATION

The goal of this section is to give a mathematical physics explanation of exponent range $0<y\leq 2$ in the empirical power law formula (2) on the ground of the Lévy stable distribution theory.

The Cauchy problem of the standard diffusion equation is expressed as (Kythe, 1996)

$$\frac{\partial p}{\partial t} - \kappa \frac{\partial^2 p}{\partial x^2} = 0, \quad (54)$$

$$p(x,0) = f(x), \qquad -\infty \prec x \prec \infty, \ t \succ 0, \quad (55)$$

where $\kappa$ represents the diffusion coefficient, and $f(x)$ is a sufficiently well-behaved function. It is well known that the solution of (54) and (55) reads

$$p(x,t) = \int_{-\infty}^{\infty} f(y) G(x-\xi,t) d\xi, \qquad (56)$$

where

$$G(x-\xi,t) = \frac{1}{\sqrt{4\pi\kappa t}} e^{-(x-\xi)^2/4\kappa t} \qquad (57)$$

is the fundamental solution (free space Green function) of this diffusion Cauchy problem, which exactly satisfies equation (54), i.e.,

$$\frac{\partial G}{\partial t} - \kappa \frac{\partial^2 G}{\partial x^2} = \delta(x-\xi, t-\tau). \qquad (58)$$

Here $\delta$ denotes the Dirac function. It is observed that

$$G(x-\xi,t) = g_G(x,\sigma) = \frac{1}{\sigma\sqrt{2\pi\kappa}} e^{-(x-\xi)^2/2\kappa\sigma^2}, \qquad (59)$$

where $\sigma^2 = 2t$. It is thus clear that such a fundamental solution can be interpreted as the Gauss normal probability density function $g_G(x,\sigma)$ with the variance $\sigma^2$ linearly proportional to time. This is a special case of the so-called $\beta$-stable Lévy distribution when $\beta=2$ (Gorenflo and Mainardi, 1998; Herrchen, 2000). Below is a definition of the stable distribution without proof.

**Definition 1**. A random variable $X$ has a stable distribution $g(x)=P\{X<x\}$ if there is a positive number $c_n$ ($n>2$) and a real number $d_n$ such that

$$X_1 + X_2 + \ldots + X_n \stackrel{d_n}{=} c_n X + d_n, \tag{60}$$

where $X_1$, $X_2$, …$X_n$ denote mutually independent random variables with identical distribution $P(x)$ of $X$. (60) means that the random variables on both sides have the same probability distribution, and then $P(x)$ is called a stable distribution.

The above definition of the Lévy stable distribution is based on the generalized central limit theorem, which characterizes the basin of attraction of the Lévy distribution (Jespersen, 1999). In other words, a distribution is stable if a sum of identically distributed variables has the same kind of distribution as the individual variables in the sum (Gorenflo and Mainardi, 1998). The concept of the stable distribution plays a fundamental role in probability theory. $c_n$ and $d_n$ in (60) are the scaling and shift coefficients. Referring to Lévy (1954) and Feller (1971),

$$c_n = n^{1/\beta}, \qquad 0 \prec \beta \leq 2, \tag{61}$$

where $\beta$ is characterized by the so-called Lévy index of stability. Using $P_\beta(x)$ instead of $P(x)$, we mean that the random variable $X$ has a stable probability distribution of the Lévy index $\beta$. The probability density function of distribution $P_\beta(x)$ for the random variable $X$ is given by

$$g(x) = \frac{dP_\beta(x)}{dx}. \tag{62}$$

If the random variable $-X$ has the same distribution as $X$, the stable distribution $P_\beta(x)$ is called symmetric. Notably, the symmetric stable distribution $P_2(x)$ ($\beta=2$) is the Gaussian distribution, and its probability density function is shown in (59) (Herrchen, 2000). In general, a symmetric multivariate Lévy $\beta$-stable probability density function is the fundamental solution of the corresponding generalized diffusion equation (also see equation (40)) (Gorenflo and Mainardi, 1998)

$$\frac{\partial p}{\partial t} - \zeta\left(-\frac{\partial^2}{\partial x^2}\right)^{\beta/2} p = 0, \qquad (63)$$

in which the second term spatial derivative in the standard diffusion equation (54) is replaced with a fractional Laplacian of order $0<\beta/2\leq 1$ (Hanyga, 2001b). By comparing (63) and (40), it is now clear that the fractional Laplacian $(-\Delta)^{y/2}$ in (40) underlies the $y$-stable Lévy process. Therefore, from a Levy statistics point of view, the empirical exponent range of the power law attenuation (see formula (2))

$$0 \prec y \leq 2 \qquad (64)$$

obeys the conditions set by a stable Lévy process in terms of the presented fractional Laplacian models (40). It is also easy to understand that the Fourier transform solution (42) of anomalous diffusion equation (40) is in fact the Fourier transform of the probability density function of the corresponding $y$-stable Lévy distribution, i.e.,

$$\hat{g}(k, y) = Ce^{-Dk^y}, \qquad (65)$$

where $D = 2\alpha_0 c_0^{1+y} t$ can be seen as a scale factor of the Lévy distribution. A $y$-stable distribution requires the power $y$ of the Fourier transform of the probability density function to be positive but not greater than 2. (65) is also interpreted as the characteristic function of a Lévy process (Hanyga, 2001a).

The above analysis shows that the empirical range limit of exponent $y$ in the power law attenuation can be derived from the basic Lévy statistics principle. Herrchen (2001) points out that there are direct connections between fractional calculus, stable distribution and self-similarity, and notes that the self-similarity extends usually only over a finite range in real physical problems. In terms of this mathematical physics background, the media having y>2 are not statistically stable in nature and the power law takes effect over a finite frequency range. This is in agreement with many experiment observations. It is noted that the Lévy process does not include y=0, which means that the media obeying absolutely frequency-independent attenuation is simply an ideal approximation.

Eq. (63) describes the anomalous diffusion (y≠2) (Jespersen, 1999) in comparison to the usual Brownian diffusion of Gaussian statistics (y=2) governed by the standard diffusion equation (54). In particular, y=1 corresponds to the Cauchy distribution whose probability density function is

$$g_C(x,\gamma) = \frac{1}{\pi} \frac{\gamma}{x^2 + \gamma^2}, \qquad (66)$$

where $\gamma$ represents the semi-interquartile range. Feller (1971) gives the probability distribution functions of all other stable Lévy distributions. It is noted that as a non-Gaussian process, the Cauchy probability density function only has finite absolute moments of order (0,1]. In general, for a y-stable Lévy process, only the absolute moments of order (0,y] are finite. In contrast, the absolute moments of any positive order of the Gauss probability density function are finite. In addition, the Gaussian normal distribution is the unique Lévy stable distribution with finite variance (Gorenflo and Mainardi, 1998). But nevertheless the probability density functions of all stable Lévy distributions are unimodal and bell-shaped (Gorenflo and Mainardi, 1998). There are also lots of reports on the so-called random walk model as a discrete form correspondence of the fractional Laplacian models, and we do not plan to discuss it here since it is beyond the scope of this paper.

## V. CONCLUDING REMARKS

Attenuation plays an essential part in many acoustics applications, for instance, the ultrasound second harmonic imaging and high intensity focused ultrasound beam for therapeutic surgery. Compared with the Szabo's time convolutional integral model of the power law attenuation, the present fractional Laplacian time-space model has a uniform and simpler expression. More importantly, most of anomalous (non-Gaussian process, $y \neq 2$) thermoviscous attenuations occur in spatially inhomogeneous environments (Henry and Wearne, 1999), notably biomaterials and geological random media, whose micro geometry mostly has fractal dimension structure in space. The power law formula (2) also shows that $y$ is independent of frequency $\omega$ (time scale). It is therefore reasonable to think that $y$ may in fact underlie the spatial fractal. For example, $y$ varies with different human body tissues, which have different spatial microstructures. We thus conclude that a spatial representation of the dissipation is physically more valid than the time representations. For the fractional derivative models with higher than the second order time derivative, it is also not a simple task to obtain the initial conditions of the second order derivative since most physical systems only provide the zero- and first-order initial conditions. Compared with the Szabo's models, numerical advantage of the fractional calculus models is that the corresponding mathematical apparatus have been intensively researched and the standard solution package could be well developed to handle different cases (Ochmann and Makarov, 1993).

We need to mention again that this study does not involve anisotropic media. It is expected that such research needs to use the so-called Feller-Lévy stable distribution which depends not only on the Lévy stability index $\beta$ ($0<\beta \leq 2$) previously discussed in section 5 but also on a skewness coefficient $\theta$ restricted by $\beta$ (Feller, 1971; Gorenflo and Mainardi, 1998). For example, the anisotropic fractional Laplacian diffusion equation discussed in Hanyga (2001a) involves the direction-dependent probability. We plan to pursue the fractional Laplacian thermoviscous models of the lossy anisotropic media in a subsequent paper.

**ACKNOWLEDGMENTS**

The work reported here is sponsored by Simula Research Laboratory with the project "Mathematical and numerical modeling of medical ultrasound wave propagation".

**APPENDIX: POWER LAW ATTENUATION, FRACTAL, AND FRACTIONAL LAPLACIAN**

Mandelbrot (1982) has already noted the connections between the Lévy stable distribution and fractals due to the inherent self-similarity of the Lévy probability density functions. An object is considered self-similar if it characterizes the same on any scale. Fractal underlies self-similarity which is described by a power law like (2), which is restated as

$$y = \frac{\ln \alpha(\omega)/\alpha_0}{\ln|\omega|}. \tag{A1}$$

(A1) reveals a self-similar property, i.e., scale invariance (Zahle, 1997). $y$ is thus interpreted as the "dimension", known as the Hausdorff dimension in the literature. In other words, the power law (2) on different frequencies (time scales) underlies the invariant parameter $y$ of the self-similarity. It is known that the $\beta$-stable Lévy distribution ($0<\beta\leq 2$) leads to self-similarity, i.e., fractal (Sato, 1999). This explains the fact that the fractal dimensions of practical interest mostly fall between 0 and 2. It is also noted from section 5 that the fundamental solution of the parabolic fractional Laplacian equation (60) exhibits self-similarity (Hanyga, 2001b). In conclusion, there are underlying connections between the Lévy index of stability, fractal dimension, exponent of power law attenuation, and the order of the fractional Laplacian in modelling of the anomalous thermoviscosity. Kolwankar and Gangal (1996) elaborate the explicit connections between self-similar fractals and the fractional derivative.

The fact that fractal *y* is mirrored by the *y/2* order fractional Laplacian discloses diffusion mechanism on fractals. The absorption behavior is found to be closely dependent on the fractal microstructures of media, while the orders of all other differential terms in the hyperbolic or parabolic model equation models are independent of fractal effects. This analysis unveils that the power coefficient *y* of the frequency-dependent attenuation can be exactly reflected by the order of the fractional Laplacian model. Varying absorption coefficient *y* over different human body tissues means the multifractal, i.e., continuous or quantum dimension variation, which are typically characterized with the fractal features of local similarity.